\documentclass[fleqn,usenatbib]{mnras}

\usepackage{newtxtext,newtxmath}
\usepackage[T1]{fontenc}  
\usepackage{ae,aecompl}

\usepackage{graphicx}	
\usepackage{float}
\usepackage{amsmath}	
\usepackage{braket}
\usepackage{cleveref}

\title[]{Superradiance and Periodic 6.7 GHz Methanol Flaring in G22.356+0.066}

\author[Rashidi et al.]{T. Rashidi,$^{1}$ V. Anari,$^{1}$ A. Bartkiewicz,$^{2}$ P. Wolak,$^{2}$ M. Szymczak,$^{2}$ F. Rajabi,$^{1}$\thanks{E-mail: rajabf1@mcmaster.ca}   
\\
$^{1}$Department of Physics and Astronomy, McMaster University, 1280 Main Street West
Hamilton, Ontario, L8S 4M1, Canada \\
$^{2}$Institute of Astronomy, Faculty of Physics, Astronomy and Informatics, Nicolaus Copernicus University, Grudziadzka 5, 87-100 Toru\'{n}, Poland \\
}

\date{}

\begin{document}
\label{firstpage}
\pagerange{\pageref{firstpage}--\pageref{lastpage}}
\maketitle

\begin{abstract}

We present a comprehensive analysis of the periodic flares observed in the 6.7 GHz methanol transition in G22.356+0.066, utilizing the Maxwell-Bloch equations (MBEs) as a framework to model these phenomena. By solving the one-dimensional MBEs, we describe the behavior of both the quasi-steady-state maser and transient superradiance regimes. Our findings indicate that the observed periodic flares, with varying timescales across different velocities, are consistent with the characteristics of Dicke's superradiance, triggered by a common radiative pump in regions of varying inverted column densities. This work provides new insights into the physical processes governing variability in maser-hosting regions and underscores the significance of superradiance as a powerful radiation mechanism in astrophysical environments.

\end{abstract}

\begin{keywords}
masers – radiation mechanisms: non-thermal – ISM: individual objects: G22.356+0.066
\end{keywords}


\section{Introduction}\label{sec:Introduction}

G22.356+0.066 (also known as G22.357+0.066 and IRAS18290-0924) is a high-mass young stellar object (HMYSO) located at a distance of \(4.3^{+3.8}_{-1.3} \, \text{kpc}\) \citep{Reid2019}. It is part of the Giant Molecular Complex G23.3-0.3, which comprises several HII regions and supernova remnants, serving as star-forming sites \citep{Zhang2023}.

HMYSOs are known to host variable methanol masers \citep{Goedhart2004, Van2009, Szymczak2012}. In 2011, \citet{Szymczak2011} reported the discovery of periodic flares of the 6.7 GHz methanol in G22.356+0.066 using the Toru\'{n} 32-m radio telescope. The flares appeared with a period of approximately 178 days (178.2 ± 1.9), with a rise-to-decay ratio of 0.34 \citep{Szymczak2015}.

The maser activity in G22.356+0.066 is not limited to 6.7 GHz methanol and is also reported in some velocity components of the 12.2 GHz methanol \citep{Breen2016}, 22 GHz water \citep{Breen2011}, and 1665 MHz OH \citep{Beuther2019} transitions. However, the flux density in these lines is either faint or the available data are too sparse, with no reports of periodicity.

A number of scenarios have been proposed to explain periodic flaring activities in maser-hosting regions, most of which are based on the maser regime of radiation. Some of these scenarios are based on the disturbance of the masing region by shock waves or clumps. Others are based on variations in the pump photons and/or the background radiation field. However, when the flux densities return to nearly the same quiescent level after flaring, this indicates that the masing region was not significantly affected by whatever mechanism caused the flare. Therefore, the disturbance scenarios involving shock waves or clumps are dismissed in such incidents \citep{Van2009} and only mechanisms involving radiative coupling with the masing region are considered viable. Examples of such mechanisms include the colliding wind binary \citep{Van2009, Van2011}, cyclic accretion onto a young binary system \citep{Araya2010}, and stellar pulsation \citep{Inayoshi2013}. 

More recently, a new regime of radiation based on Dicke's superradiance has been introduced in maser-hosting regions to model periodic \citep{Rajabi2023,Houde2024} and non-periodic \citep{Rajabi2016a, Rajabi2017, Rajabi2019} flaring events. In essence, superradiance is a transient, highly efficient radiation mechanism that manifests itself through sharp rises in flux densities, with burst durations scaling with some fundamental properties of the transition, such as the wavelength of the transition, and physical parameters characterizing the source, such as the inverted column density. In \citet{Rajabi2023}, superradiance provides a natural explanation for periodic flaring in G9.62+0.20E at four different transitions (6.7 GHz and 12.2 GHz methanol lines, as well as OH 1665 MHz and 1667 MHz) exhibiting different timescales in response to a common radiative pump. In \citet{Szymczak2011}, it is noted that the profiles of periodic methanol flares in G9.62+0.20E and G22.356+0.066 are very similar. This suggests that the same underlying physical mechanism could be responsible for the periodicity in both sources. It is, therefore, a natural step to extend the superradiance analysis performed on G9.62+0.20E to G22.356+0.066. However, in G22.356+0.066, the periodicity is only reported in the 6.7 GHz methanol transition line, which will therefore be the focus of this paper. 

This paper is structured as follows: In Sec.~\ref{sec:SRvsMaser}, we briefly discuss the two distinct regimes of radiation, namely the maser action and superradiance. Sec.~\ref{sec:AnalysisModel} provides a detailed discussion of the modelling and analysis. In Sec.~\ref{sec:discussion}, we evaluate the validity of the superradiance model presented and conclude the paper. The MBEs used in our analysis are detailed in Appendix~\ref{subsec:MBEs}. Finally, the autocorrelation analysis developed to quantify and identify variations in the timescales of flares across different velocity components is summarized in Appendix~\ref{subsec:autocorrelation}.

\section{The maser and superradiance regimes}\label{sec:SRvsMaser}  

As discussed extensively in \citet{Rajabi2020} and \citet{Rajabi2023}, the Maxwell-Bloch equations (MBEs) (see Appendix~\ref{subsec:MBEs}) provide a comprehensive description of a compound system comprising an extended group of $N$ two-level molecules (or atoms) and a radiation field. This set of equations tracks the underlying difference in population between the energy levels, the induced polarization in the system, and the evolution of the radiation field \citep{MacGillivray1976, Gross1982, Benedict1996}. A detailed study of the MBEs for a gas with an initial population inversion reveals two complementary regimes: the quasi-steady-state maser regime, where the stimulated emission process dominates, and the fast transient superradiance regime \citep{Feld1980,Rajabi2020, Rajabi2023, Houde2024}.

In the quasi-steady-state maser regime, temporal variations in the population inversion, polarization, and radiation field occur on an evolution timescale (\(T_{\mathrm{e}}\)) longer than the noncoherent decaying and dephasing timescales (\(T_1\) and \(T_2\)) that characterize the system's response. Consequently, a maser system in this regime can quickly adjust to the excitation signal. For example, a saturated maser's response to an inversion pump closely tracks the profile of the pump itself \citep{Rajabi2023}.

In the fast transient superradiance regime, the system responds differently to variations in the pump signal whenever \(T_{\mathrm{e}} \approx T_{\mathrm{R}} < T_1, T_2\), with \(T_{\mathrm{R}}\) being the characteristic superradiance timescale. The transient intensity triggered by the excitation is thus a characteristic response of the system and is largely independent of the shape of the input signal, whether it is a pump or a trigger \citep{Rajabi2023}. Another distinctive aspect of the superradiance is that its peak intensity scales with the square of the number of molecules involved (\(N^2\)), in contrast to the linear scaling (\(N\)) observed in a saturated maser. The characteristics of the transient response, such as its duration, profile, and strength, are influenced by the physical conditions of the environment, including the relative significance of \(T_1\) and \(T_2\) compared to \(T_{\mathrm{R}}\), as well as the inverted column density. Additionally, the parameters that define the radiative transition, such as the wavelength and the Einstein spontaneous emission coefficient, play a crucial role. More precisely, the characteristic superradiance timescale is defined as:

\begin{equation}\label{eq:TR}
T_{\mathrm{R}} = \frac{8\pi}{3 n L \lambda^2}\tau_{\mathrm{sp}},
\end{equation}
where \(\tau_{\mathrm{sp}} = \Gamma^{-1}\) is the spontaneous emission timescale of the transition, \(\lambda\) is the wavelength of radiation, \(n\) is the population inversion density, and \(L\) is the length of the system (\(nL\) is the inverted column density) \footnote{Note that we are focusing on radiating systems with a cylindrical geometry.}. It should be noted that while \( T_{\mathrm{R}} \) can be interpreted simply in single superradiance burst events, it is more complex to define in the case of periodic or multi-flaring events. We thus define an initial \( T_{\mathrm{R,0}} \) before the onset of the first modelled flare, using the initial inverted column density to establish a relevant timescale for interpreting the MBEs solutions (see Sec.~5 of \citealt{Rajabi2023}).

As we will show in Sec.~\ref{sec:results}, the observational data of the periodic flaring in G22.356+0.066 support the superradiance model. We observe that the speed at which a superradiance system responds to an excitation varies among different velocity components. This is quantified through an autocorrelation analysis (see Appendix~\ref{subsec:autocorrelation}), which reveals the presence of different timescales at various velocities. This is a phenomenon that can be readily explained within the context of superradiance \citep{Rajabi2017, Rajabi2023}.

More precisely, this implies that, when a medium is pumped, different velocity components of the 6.7 GHz methanol transition might couple differently to the same pumping source resulting in an inversion density \(n\) that can vary throughout the medium.

\section{Analysis and model}\label{sec:AnalysisModel}

\subsection{Data}\label{subsec:data}

The data sets used in our analysis were obtained with the Toru\'{n} 32-m telescope between June 2009 and April 2014 at irregular intervals; see \citet{Szymczak2015} for more details. The observations revealed cyclic variations in flux density for features within the velocity range of 78.95~km~s$^{-1}$ to 83.50~km~s$^{-1}$, which are the focus of this investigation. The periodic features in G22.356+0.066 consist of a non-zero quiescent level, upon which the periodic flares are superimposed. In Appendix~\ref{subsec:autocorrelation}, we selected only those features with a nearly constant quiescent level for modelling.

\subsection{Model and results}\label{sec:results}

To model the flares in G22.356+0.066, similar to \citet{Rajabi2019} and \citet{Rajabi2023}, we solve the MBEs using the fourth-order Runge-Kutta method for a one-dimensional sample with an inverted column density \( nL \) \citep{Mathews2017}. To simulate periodic flaring, we employ a periodic pump:

\begin{equation} \label{eq:pump}
    \Lambda_N(z, \tau) = \Lambda_0 + \sum_{m=0}^{\infty} \frac{\Lambda_{1,m}}{\cosh^2{\left[ \left( \tau - \tau_0 - m\tau_1\right) / T_p \right]}}.
\end{equation}
This pump propagates along the symmetry axis ($z$-axis) of the sample. It consists of a constant term \( \Lambda_0 \), modelling  quiescent flux levels, and a sequence of pump pulses with amplitudes \( \Lambda_{1,m} \), which can vary from pulse to pulse. The period of the pulses, \( \tau_1 \), is set to 178.5 days, closely matching the reported period of the flares at the source. The duration of the pump pulses is determined by \( T_p \), and a time delay \( \tau_0 \) is chosen to fit superradiance to the data. 

\begin{figure}
    \centering
    \includegraphics[width=\linewidth, keepaspectratio]{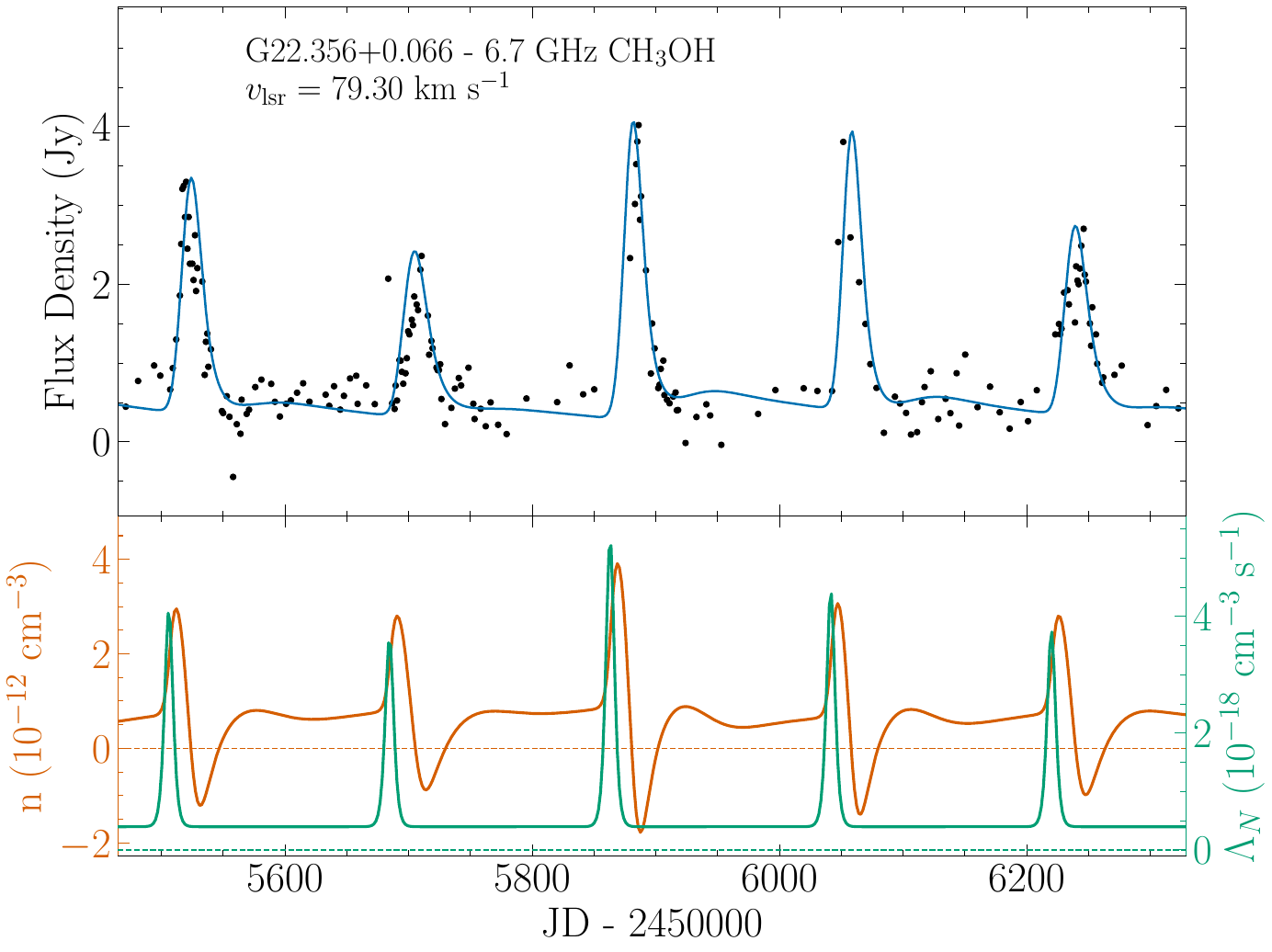}
    \caption{Top: Periodic flares of the 6.7 GHz methanol transition line in G22.356+0.066 for \( v_{\mathrm{lsr}} = 79.30 \) km s\(^{-1}\) (black dots) and the flux density yielded from MBEs (solid blue curve) are plotted as a function of time in Julian Day-2450000. Bottom: The selected periodic pump with \( \Lambda_0 = 3.97 \times 10^{-19} \) cm\(^{-3}\) s\(^{-1}\) and \( \Lambda_i = 4.45 \times 10^{-18} \) cm\(^{-3}\) s\(^{-1}\), along with pump amplitudes \( a_m \) of 0.84, 0.73, 1.10, 0.91, and 0.76 (green), respectively, and the population inversion density (\( n \)) response to the pump (vermilion). The model yields an initial inverted column density of \( n_0 L = 5.28 \times 10^3 \) cm\(^{-2}\), corresponding to \( T_{\mathrm{R},0} = 5.03 \times 10^4 \) s.}
    \label{fig:SR-79.30}
\end{figure}

\begin{figure}
    \centering
    \includegraphics[width=\linewidth, keepaspectratio]{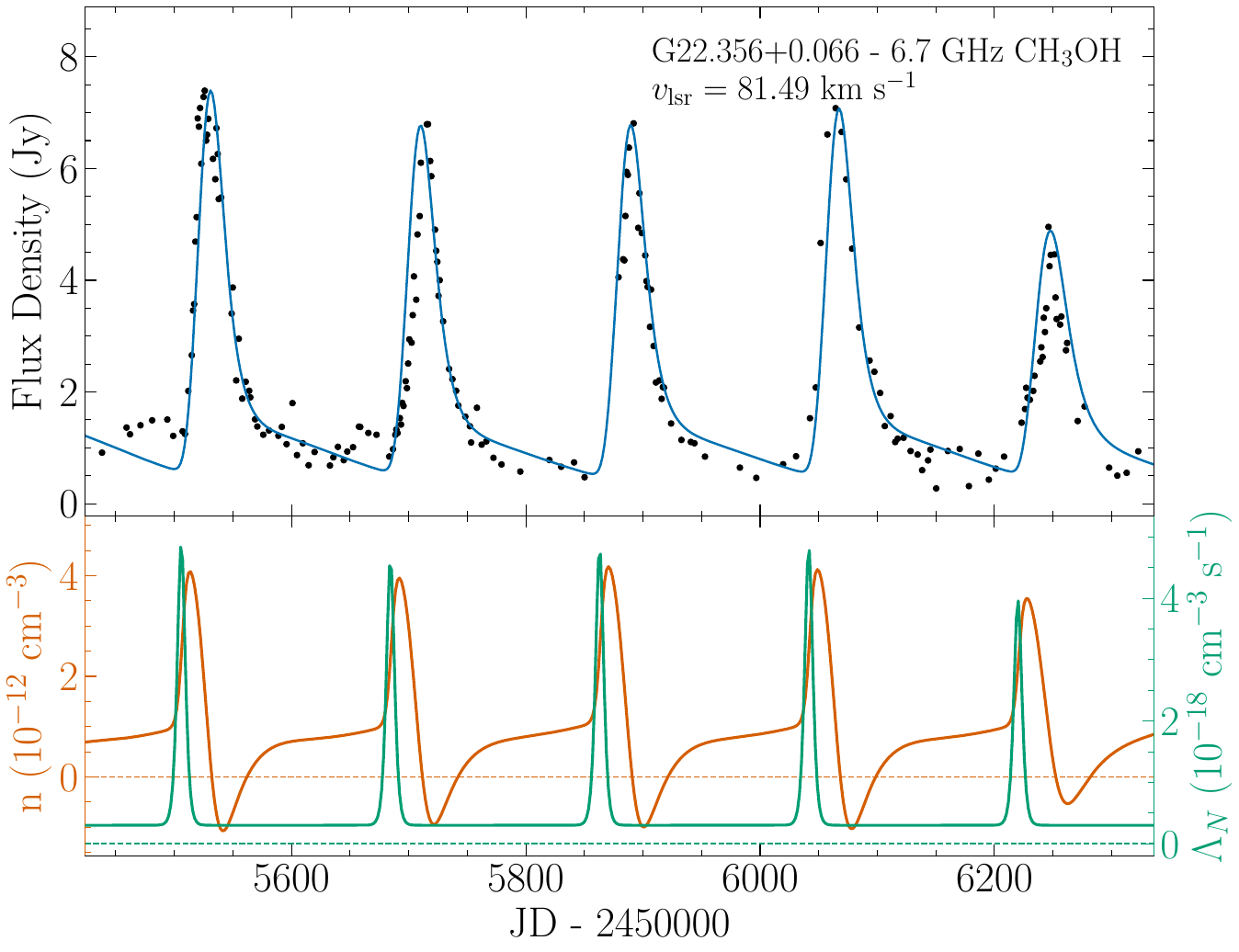}
    \caption{Top: Same as Figure~\(\ref{fig:SR-79.30}\), but for \( v_\mathrm{lsr} = 81.49 \)~km~s\(^{-1}\). 
Bottom: Same as Figure~\(\ref{fig:SR-79.30}\), with \( \Lambda_0 = 3.01 \times 10^{-19} \)~cm\(^{-3}\)~s\(^{-1}\) and \( \Lambda_i = 4.62 \times 10^{-18} \)~cm\(^{-3}\)~s\(^{-1}\), and pump amplitudes \( a_m \) of 1.00, 0.94, 0.98, 0.98, and 0.80, respectively. The model yields \( n_0 L = 4.00 \times 10^3 \)~cm\(^{-2}\), corresponding to \( T_{\mathrm{R},0} = 6.63 \times 10^4 \)~s.}
    \label{fig:SR-81.49}
\end{figure}

The results of our simulations for four selected velocities are presented in Figures~\(\ref{fig:SR-79.30}\)–\(\ref{fig:SR-81.49}\). The black dots in the top panels represent the observational data, while the solid blue curves correspond to the flux density obtained by solving the MBEs. The bottom panels display the population inversion (in vermilion) and the periodic pump (in green), with the horizontal axis showing time in Julian date.

In these figures, after each pump pulse, the population inversion increases, followed by a decay, with the peaks of the population inversion occurring with a delay relative to the peaks of the inversion pump. It is also seen that the peaks of the inversion pump are narrower than those of the population inversion. 

To fit the superradiance models to the observed data, we varied \( \Lambda_0 \) and \( \Lambda_{1,m} \) in equation~(\ref{eq:pump}). We kept the parameters \( T_1 = 1.77 \times 10^7 \, \text{s} \) (\(\simeq 205\) days), \( T_2 = 1.02 \times 10^6 \, \text{s} \) (\(\simeq 11.8\) days), \( \tau_1 = 1.52 \times 10^7 \, \text{s} \) (\(\simeq 178.5\) days), and \( T_p = 3.89 \times 10^5 \, \text{s} \) (\(\simeq 4.5\) days) constant across all velocities. To determine \( \Lambda_{1,m} \), we used a constant \( \Lambda_i \) for each velocity and introduced a varying amplitude \( 0 < a_m\leq 1 \) for each flare, defining \( \Lambda_{1,m} = a_m \Lambda_i \).

Due to the non-linear nature of the MBEs and the fact that the response of the superradiant system does not precisely follow the pump (see Sec.~\ref{sec:SRvsMaser}), the choice of parameters \( \Lambda_0 \), \( \Lambda_i \), and \( a_m \) is not unique. Specific values for these parameters are provided in the figure captions. For each velocity component, the initial inversion density \( n_0 \) is calculated as \( n_0 = \Lambda_0 T_1 \), which in all cases exceeds the threshold needed for superradiance and the ensuing saturated maser regime \citep{Rajabi2020, Rajabi2023}. Thus each model yields an initial inverted column density \( n_0L \), corresponding to an initial characteristic superradiance timescale \( T_{\mathrm{R,0}} \), as defined by equation~(\ref{eq:TR}). This timescale is determined before the onset of the first superradiance burst, when only very weak levels of radiation are present in the simulations. It is important to note that \( T_{\mathrm{R,0}} \ll T_1, T_2 \) in all cases, suggesting the potential for transient superradiance to follow.

\begin{figure}
    \centering
    \includegraphics[width=\linewidth, keepaspectratio]{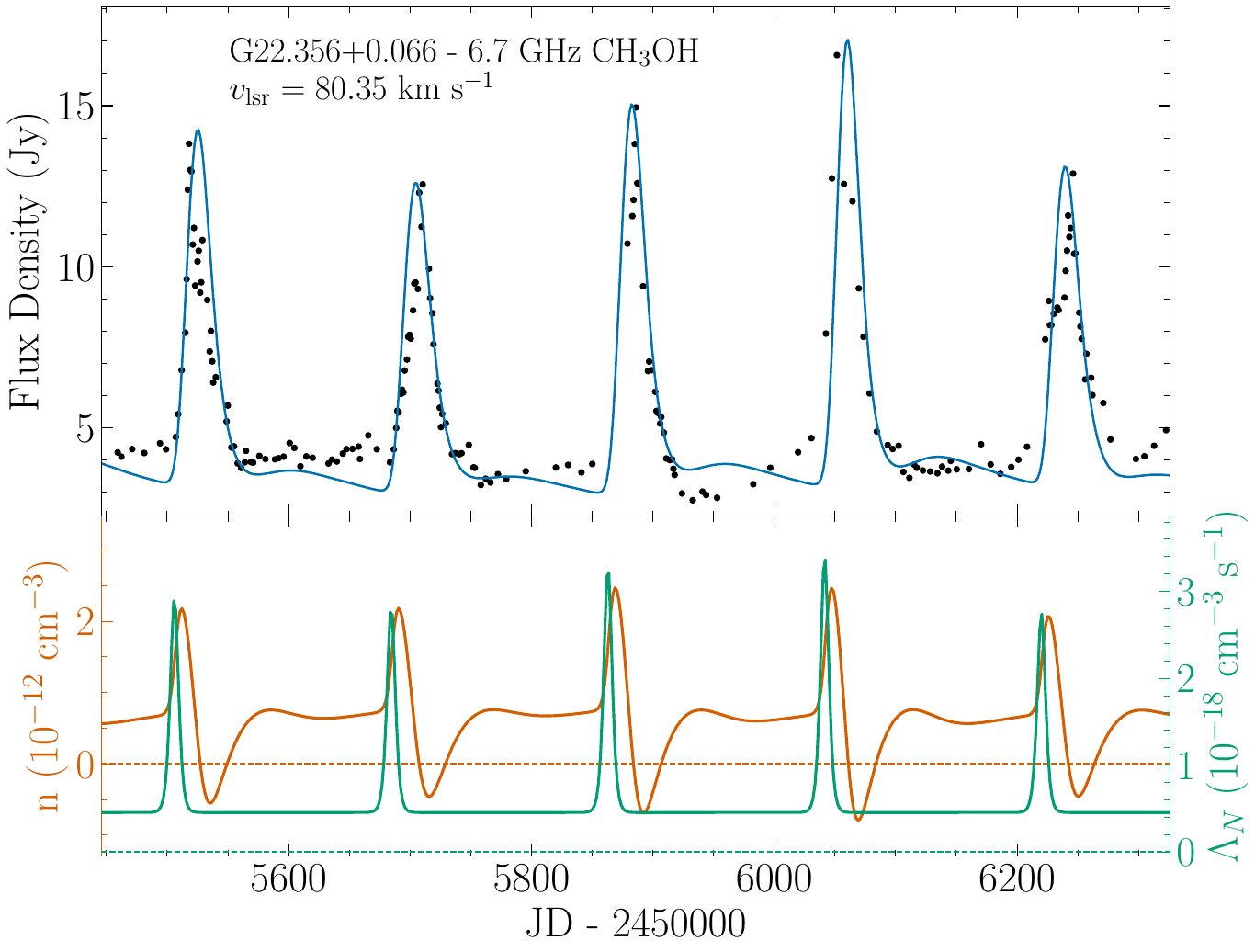}
    \caption{Top: Same as Figure~\(\ref{fig:SR-79.30}\), but for \( v_\mathrm{lsr} = 80.35 \)~km~s\(^{-1}\).
Bottom: Same as Figure~\(\ref{fig:SR-79.30}\), with \( \Lambda_0 = 4.53 \times 10^{-19} \)~cm\(^{-3}\)~s\(^{-1}\) and \( \Lambda_i = 2.93 \times 10^{-18} \)~cm\(^{-3}\)~s\(^{-1}\), and pump amplitudes \( a_m \) of 0.84, 0.81, 0.97, 1.00, and 0.79, respectively. The model yields an initial inverted column density of \( n_0 L = 6.02 \times 10^3 \)~cm\(^{-2}\), corresponding to \( T_{\mathrm{R},0} = 4.41 \times 10^4 \)~s.}
    \label{fig:SR-80.35}
\end{figure}
\begin{figure}
    \centering
    \includegraphics[width=\linewidth, keepaspectratio]{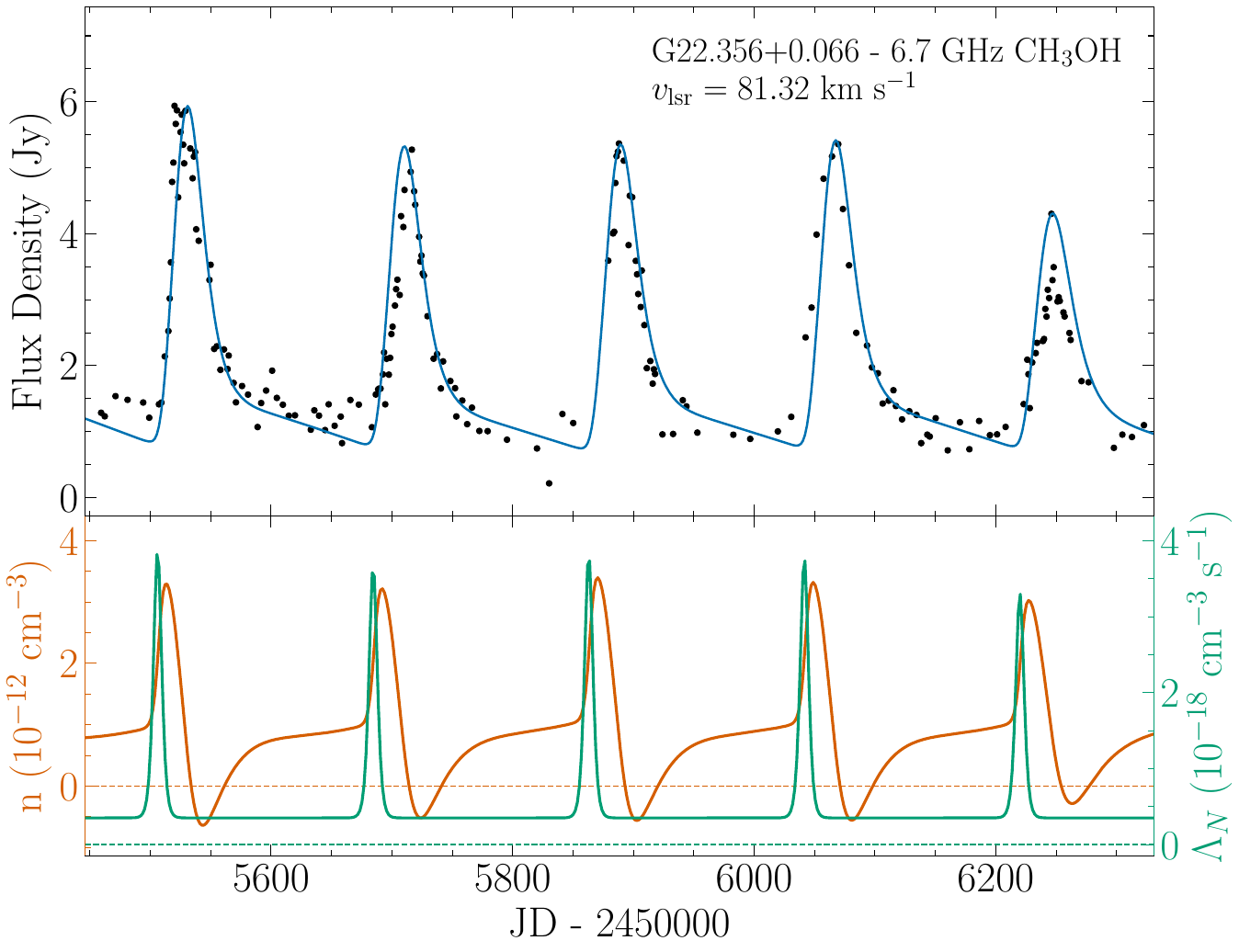}
    \caption{Top: Same as Figure~\(\ref{fig:SR-79.30}\), but for \( v_\mathrm{lsr} = 81.32 \)~km~s\(^{-1}\). 
Bottom: Same as Figure~\(\ref{fig:SR-79.30}\), with \( \Lambda_0 = 3.51 \times 10^{-19} \)~cm\(^{-3}\)~s\(^{-1}\) and \( \Lambda_i = 3.62 \times 10^{-18} \)~cm\(^{-3}\)~s\(^{-1}\), and pump amplitudes \( a_m \) of 0.97, 0.92, 0.96, 0.95, and 0.82, respectively. The model yields \( n_0 L = 4.66 \times 10^3 \)~cm\(^{-2}\), corresponding to \( T_{\mathrm{R},0} = 5.69 \times 10^4 \)~s.}
    \label{fig:SR-81.32}
\end{figure}

\begin{figure}
    \centering
    \includegraphics[width=0.8\linewidth, keepaspectratio]{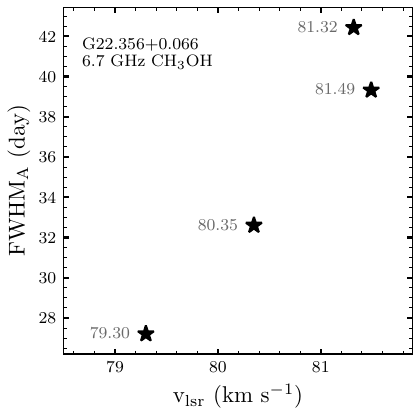}
    \caption{Full width at half maximum (FWHM) of the central peak of the autocorrelation curve for four flaring features in G22.356+0.066. The vertical axis shows the FWHM$_{\mathrm{A}}$ in days, and the horizontal axis shows the velocity of the feature in~km~s$^{-1}$. The velocity corresponding to each FWHM$_{\mathrm{A}}$ is explicitly labeled on the graph.}
    \label{fig:FWHM}
\end{figure}
As discussed in Appendix~\ref{subsec:autocorrelation}, we used an autocorrelation analysis to compare the flare duration for different velocities. The width of the central peak in the autocorrelation of a periodic function is a measure of the average widths of the flares. These widths, FWHM$_{\text{A}}$, are compared in Figure~\ref{fig:FWHM}, where the vertical axis represents the FWHM$_{\text{A}}$ in days and the horizontal axis shows velocity in~km~s$^{-1}$.

\section{Discussion}\label{sec:discussion}

Our analysis demonstrates that our model based on the MBEs effectively captures and reproduces the duration of flare profiles of the 6.7 GHz methanol maser at four distinct velocity components in G22.356+0.066. A key feature of the model is its ability to use a common pump across different velocity components, with only the pump amplitudes being modified. Specifically, while the pump-related timescales $\tau_1$ and $T_p$ were held constant, the pump parameters $\Lambda_0$ and $\Lambda_{1,m}$ were adjusted to fit the model to the observed data. These adjustments in pump amplitude successfully reproduced the varying flare durations, which show a notable increase from $v_\mathrm{lsr}=79.30$~km~s$^{-1}$ to $v_\mathrm{lsr}=81.49$~km~s$^{-1}$, as quantified by autocorrelation analysis and illustrated in Figure~\ref{fig:FWHM}. The variations in pump amplitude can be interpreted as differences in the coupling of the pump to the systems responsible for the flares, potentially due to factors such as orientation, sample geometry, and other local conditions \citep{Rajabi2023}.

It is commonly understood that 6.7 GHz methanol masers are driven by infrared (IR) radiation through radiative pumping \citep{Sobolev1997}. A nearby variable IR source within the flaring region of G22.356+0.066 was cataloged by \citet{Lucas2008}. \citet{Szymczak2015} mapped the 6.7 GHz methanol flares alongside this IR source, highlighting its location a few hundred AU east of the periodic flaring region. Although IR light curves are not available, the superradiance flares are not sensitive to the exact profile of the pumps used (e.g., symmetric or asymmetric), and similar results can be obtained with different pump profiles while maintaining a constant flare duration \citep{Houde2024}. The key point is that, within such a small physical proximity of the flares to the pump, the variable flare durations cannot be explained by traditional maser theory, where maser flares are expected to closely mimic changes in the pump for saturated masers. A common pump cannot account for the presence of substantial differences (a variation of more than 50\%) in flare timescale. However, superradiance highlights a transient regime of radiation where the flare duration, even in response to the same pump duration, can vary depending on the physical conditions realized at the source.

The maps of 6.7 GHz methanol flares in G22.356+0.066 presented by \citet{Bartkiewicz2009} and \citet{Szymczak2015} indicate that periodic flaring sources at different velocities are concentrated within a compact region of about 80~AU. Consequently, to fit the MBEs model to the light curves of various velocity channels, the dephasing and relaxation timescales, \( T_1 \) and \( T_2 \), were kept constant, implying a similar environment across the source.

Remarkably, the timescales  \( T_1 \simeq 205\) days and \( T_2 \simeq 11.8\) days, derived here closely align with those reported by \citet{Rajabi2023} for modelling the periodic flares in G9.62+0.20E, further suggesting comparable environmental conditions between the two sources. Finally, G22.356+0.066 is the second example of periodic flaring, alongside G9.62+0.20E, where we successfully model flares of different durations originating from a close physical vicinity, using the MBE framework and the superradiance mechanism.
\appendix

\section{Maxwell-Bloch Equations}\label{subsec:MBEs}

The models are derived from the MBEs \citep{Gross1982, Benedict1996, Rajabi2023}:

\begin{equation} \label{eq:MBE-n}
    \frac{\partial n'}{\partial \tau} = \frac{i}{\hbar} \left( P_0^+E_0^+ - P_0^-E_0^- \right) - \frac{n'}{T_1} + \Lambda_N
\end{equation}

\begin{equation} \label{eq:MBE-P}
    \frac{\partial P_0^+}{\partial \tau} = \frac{2i d^2}{\hbar} E_0^-n' - \frac{P_0^+}{T_2} + \Lambda_P
\end{equation}

\begin{equation} \label{eq:MBE-E}
    \frac{\partial E_0^+}{\partial z} = \frac{i \omega_0}{2 \epsilon_0 c} P_0^-,
\end{equation}
where the slow-varying-envelope and rotating wave approximations are applied. Here, \( \tau \equiv t - z/c \) represents the retarded time. The variable \( n' \) denotes half of the inverted population density \((n=2n')\), while \( P_0^+ \) and \( E_0^+ \) are the amplitudes of the polarization and electric field, respectively. 

To account for the inversion pump, the term \( \Lambda_N \), as defined in equation~(\ref{eq:pump}), is included in equation~(\ref{eq:MBE-n}). Internal fluctuations in \( n' \) and \( P_0^+ \) are introduced to initiate the temporal evolution of the system. To represent these internal fluctuations in polarization, a phenomenological polarization pump term \( \Lambda_P \) is added to equation~(\ref{eq:MBE-P}). Non-coherent relaxation and polarization dephasing are modelled by the decay terms \( -n'/T_1 \) and \( -P_0^+/T_2 \), respectively. The polarization and electric field vectors are:

\begin{equation}
    \boldsymbol{P}^{\pm}(\tau, z) = P_0^{\pm}(\tau, z) e^{\pm i \omega_0 \tau} \boldsymbol{\epsilon}_d
\end{equation}

\begin{equation}
    \boldsymbol{E}^{\pm}(\tau, z) = E_0^{\pm}(\tau, z) e^{\mp i \omega_0 \tau} \boldsymbol{\epsilon}_d,
\end{equation}
where $\boldsymbol{\epsilon}_d = \boldsymbol{d}/d$ is the unit vector for the dipole of the molecular transition, $\omega_0 = ck$ is the transition angular frequency, and $d = |\boldsymbol{d}|$ is the transition electric dipole moment. The superscripts $+$ and $-$ denote transitions between the lower and upper levels and vice versa, respectively.

\section{Autocorrelation}\label{subsec:autocorrelation}

To quantitatively assess the differences in flare durations across features at different velocities, we perform an autocorrelation analysis of the flares. The width of the central peak in the autocorrelation function provides a measure of the mean flare duration. By comparing the full widths at half maximum (FWHM$_{\mathrm{A}}$) of the autocorrelation functions for different velocities, we can demonstrate that the flaring features exhibit varying durations.

\begin{figure}
    \centering
    \includegraphics[width=\linewidth, keepaspectratio]{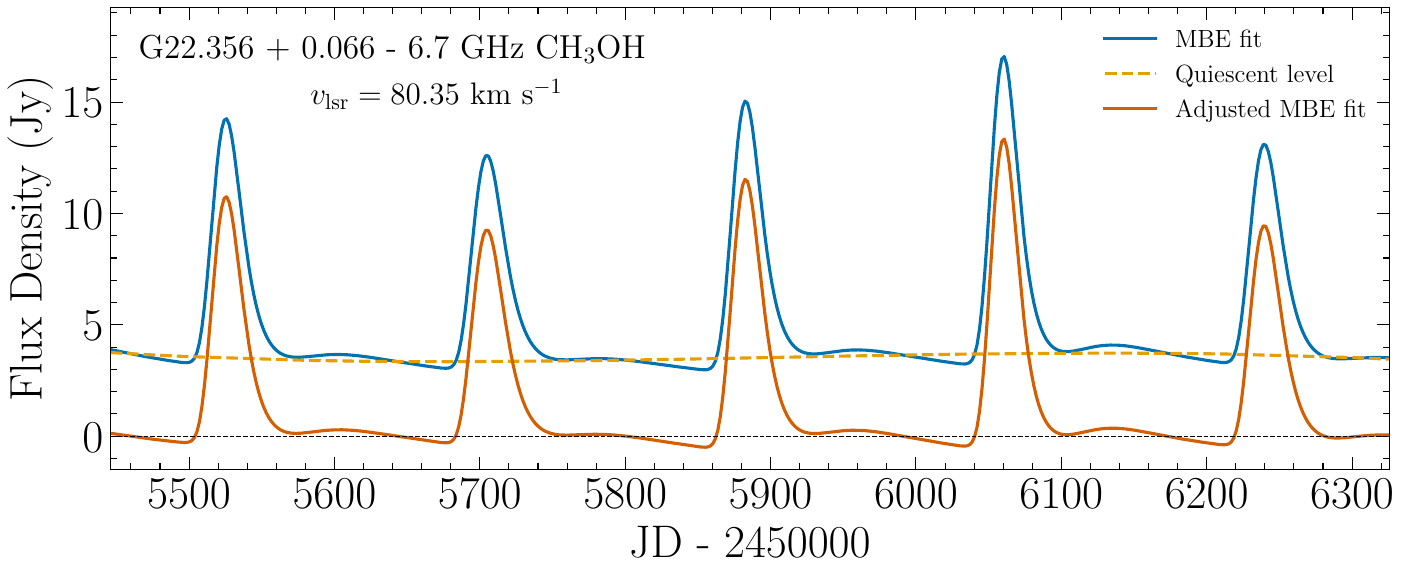}
    \caption{Removal of the non-zero quiescent level from the SR fit for the \( v = 80.35 \) km s\(^{-1}\) velocity channel in G22.356+0.066. The vertical axis shows the flux density in Jy, while the horizontal axis is for time in Julian Day \( - \, 2450000 \). The solid blue curve illustrates the fit produced by the MBEs. The dashed orange line shows a third-order polynomial fit to the quiescent level. The solid vermilion curve displays the adjusted MBEs fit, which is obtained by subtracting the third-order polynomial fit from the original MBEs fit.}
    \label{fig:baseline_80.35}
\end{figure}

\begin{figure}
    \centering
    \includegraphics[width=\linewidth, keepaspectratio]{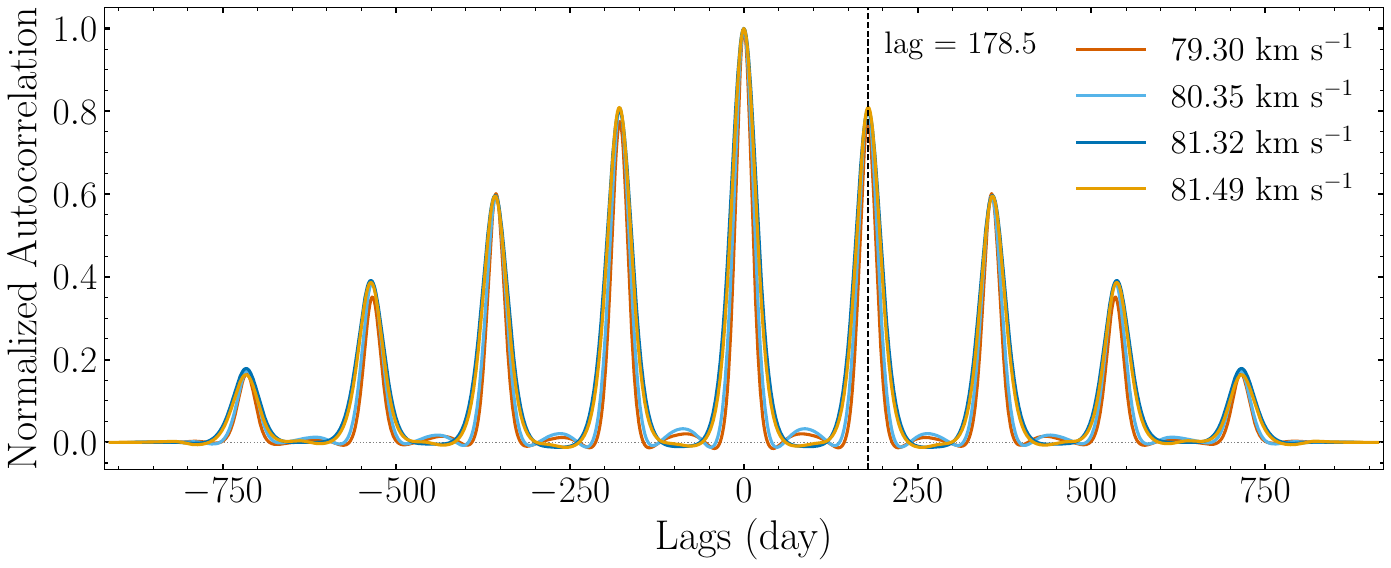}
    \caption{Autocorrelation curves for flaring features in G22.356+0.066 at velocities 79.30, 80.35, 81.32, and 81.49~km~s$^{-1}$. The vertical axis shows the normalized autocorrelation, and the horizontal axis represents lags in days. The vertical dashed line marks a lag of 178.5 day, the period of the flaring events.}
    \label{fig:autocorrelation}
\end{figure}

To obtain smooth autocorrelation curves, we use fits to the data yielded by MBEs (solid blue curves in Figs.~\ref{fig:SR-79.30} to \ref{fig:SR-81.49}), as the data are limited and unevenly spaced. We first subtract a third-order polynomial function, which fits the quiescent maser level between flares, from our models, as shown in Fig.~\ref{fig:baseline_80.35}. We then compute the normalized autocorrelation using Python's \textit{scipy.signal.correlate} function in \textit{full} mode. The resulting autocorrelation is depicted in Fig.~\ref{fig:autocorrelation}, where the vertical axis shows the normalized autocorrelation, and the horizontal axis indicates lags in days. The vertical distance between consecutive peaks is 178.5 days, corresponding to the assumed flare period in our model.

The FWHM of the central peak for each autocorrelation curve is measured and reported in Fig.~\ref{fig:FWHM}.

\section*{Acknowledgements}
F.R.'s research is supported by the Natural Sciences and Engineering Research Council of Canada (NSERC) Discovery Grant RGPIN-2024-06346. AB, PW, MS acknowledge support from the National Science Centre, Poland through grant 2021/43/B/ST9/02008. We also appreciate the valuable feedback provided by Prof. Martin Houde.

\bibliographystyle{mnras}
\bibliography{references}

\label{lastpage}

\end{document}